\documentclass[11pt]{article}
\newsavebox{\PSLASH}
\sbox{\PSLASH}{$p$\hspace{-1.8mm}/}

\input{amssym}
\begin{document}
\title{Finite Size Scaling and Conformal Curves}
\author{M. A. Rajabpour\footnote{e-mail: rajabpour@mehr.sharif.edu}
, S. Moghimi-Araghi \footnote{e-mail: samanmi@sharif.edu} \\ \\
Department of Physics, Sharif University of Technology,\\ Tehran,
P. O. Box: 11365-9161, Iran} \maketitle

\begin{abstract}
In this letter we investigate the finite size scaling effect on
SLE($\kappa,\rho$) and boundary conformal field theories and find
the effect of fixing some boundary conditions on the free energy
per length of SLE($\kappa,\rho$). As an application, we will
derive the entanglement entropy of quantum systems in critical
regime in presence of boundary operators.
  \vspace{5mm}%
\end{abstract}

\section{introduction}
Conformal field theories have found many applications in
classification of phase transitions and critical phenomena in two
dimensions. In particular the minimal models introduced in
\cite{BPZ} reveal many exact solutions to various two dimensional
phase transitions like Ising model at critical point or Pott's
model and so on \cite{difran}. These models were first considered
on the whole plane, but as many surface phenomena are very
interesting to analyze, boundary conformal field theory was soon
developed \cite{cardy84}. Essentially it was shown that conformal
field theory in the half plane with proper boundary condition
could be mapped to a whole-plane conformal field theory with just
holomorphic part.

On the other hand, recently a new method to investigate the so
called geometrical phase transitions has been developed, which
were previously described by conformal field theories. The new
method, Stochastic Loewner Evolution (SLE) \cite{Schramm} is a
probabilistic approach to study scaling behavior of geometrical
models. SLE's, which are characterized by a parameter $\kappa$,
can be simply stated as conformally covariant processes, defined
on the upper half plane, which describe the evolution of random
domains, called SLE hulls. These random domains represent critical
clusters. The idea of SLE was first developed by Schramm
\cite{Schramm}. He showed that under assumption of conformal
invariance, the scaling limit of loop erased  random walks is
SLE$_2$. Since such problems can be investigated either by using
SLE methods or by exploiting conformal field theories, there
should be a direct relation between the two, such relation was
found in \cite{bb} and later in \cite{RF1}.

More recently a generalization of these theories has emerged: the
so called SLE($\kappa,\rho$) \cite{lsw,dub}. The new theories
describe random growing interfaces in a planar domain which have
markovian property and conformal invariancy. In fact
SLE($\kappa,\rho$) is the minimal way to generalize the original
SLE while keeping self-similarity and markov property. These
models are related to conformal field theory, too
\cite{cardy04,Kyt05}.

The problem which has not been addressed yet, is the effect of
finite size scaling on SLE($\kappa,\rho$) which is in fact the
problem of finite size effect on boundary conformal field theory.
The problem could appear if you are investigating a system in
which the size of the SLE hulls are comparable with the size of
the whole system, and this happens in most of physical cases. Also
it could be used to calculate the quantum entanglement entropy for
quantum systems in $1+1$ dimensions, which is a very interesting
problem in quantum computation.

In the next section we will briefly recall the properties of
SLE$(\kappa,\rho)$ and its relation to conformal field theory, and
in the third section we'll derive the finite size effects and its
 in different problems.

\section{SLE$(\kappa,\rho)$ and its properties}

As mentioned above, SLE($\kappa,\rho$) describes random growing
interfaces in a planar domain which have markovian property and
conformal invariancy. Schramm-Loewner equation reveals the
evolution of conformal map which transforms the remaining part of
the upper half plane to the whole upper half plane. The modified
version of the equation depends on a series of numbers $\rho_j$ in
addition to the points $x_j$. In the CFT picture, at these point
the proper boundary changing operators are inserted
\cite{cardy04}. In fact using the SLE($\kappa,\rho)$ equation, one
is able to show that there should be time independent states
$|h,x_j\rangle$, which should annihilate if a certain combination
of generators of Virasoro algebra is applied on them. It is then
shown in \cite{cardy04} that the operators associated with jumps
in piecewise Drichlet boundary condition of a coulomb gas theory
can create these states. In \cite{mrr} we have generalize the
argument to the other boundary conditions in coulomb gas model.
Also the effect of existence of a charge at infinity is discussed.
To be more specific we will briefly recall the method used in
\cite{cardy04,mrr}

Considering the coulomb gas action $S=(4\pi)^{-1}\int
[(\partial\varphi)^2+ 2\, Q\, R\, \varphi]$, one can derive the
classical solution of equation of motion which satisfies the
boundary condition of the problem. As an example, in Neuman
boundary condition (NBC) the solution has the form
$\varphi_{cl}=\sum \lambda_{j} \ln(z-x_{j})(\bar{z}-x_{j})$ where
$\lambda_j$ are constants coming from the specific choice of the
boundary condition. Then one can consider fluctuations around this
solution and find the partition function associated with
$\varphi_{cl}$ which we call it $Z_{cl}$. This partition function
is considered to be the definition of boundary changing operators
via defining their correlation functions:
$\langle\prod_{j}\phi_{\lambda_{j}}(x_{j})\rangle ={Z_{cl}}$. Also
any expectation value $\langle O\rangle$ in presence of this
boundary condition can be defined in the following way:
\begin{equation}\label{correlation}
\langle O\rangle =\frac{\langle
O\prod_{j}\phi_{\lambda_{j}}(x_{j})\rangle
}{\langle\prod_{j}\phi_{\lambda_{j}}(x_{j})\rangle}.
\end{equation}
In particular the expectation value of energy-momentum tensor $T$
can be computed to be
\begin{equation}\label{T}
\langle T\rangle =-
(\partial\varphi_{cl})^{2}+2Q\partial^{2}\varphi_{cl}=
\sum_{j,k}\frac{\lambda_{j}\lambda_{k}}{(z-x_{j})(z-x_{k})}+2 Q
\sum_{j}\frac{\lambda_{j}}{(z-x_{j})^{2}}.
\end{equation}
OPE of this energy momentum tensors with the boundary fields
$\varphi_{\lambda}$ could be calculated to derive the weight of
these fields $h_\lambda= \lambda(\lambda+2\,Q)$.

Using the above equations,  the effects of Virasoro generators on
the boundary operators could be calculate and one observes that
they satisfy the relation coming from SLE part of the problem if
we impose the condition
$Q=\displaystyle{\frac{\lambda(-\kappa+4)}{4}}$. So the relation
between SLE($\kappa,\rho$) and boundary CFT is derived,
SLE($\kappa,\rho$) martingailes are on one side and boundary
conformal field theory is on the other. With this equivalency one
can use CFT methods to compute the effect of finite size scaling
on SLE$(\kappa,\rho)$.

\section{Finite Size Scaling}

To investigate the finite size scaling effect on
SLE($\kappa,\rho$), we consider a generic conformal field theory
living on the half complex plane. Applying the transformation $z
\rightarrow w=\frac{L}{\pi}\ln z$, the half plane is  mapped to a
strip of length $L$. Here $w$ and $z$ are the holomorphic
coordinates on the strip and the upper half plane, respectively.
The energy-momentum tensor on the strip is found to be
\begin{equation}\label{t transform}
T_{strip}(w)=\left(\frac{\pi}{L}\right)^{2}\left({T_{H}(z)z^{2}-\frac{c}{24}}\right),
\end{equation}
where $T_H$ is the energy-momentum tensor of the half plane and
$c=\frac{(6-\kappa)(3\kappa-8)}{2\kappa}$ is the central charge of
the theory and is related to a soft breaking of conformal symmetry
by introducing a microscopic scale in the system. In our case, the
expectation value of energy-momentum tensor is given by equation
(\ref{T}), so in the strip geometry, the vacuum energy density is
found to be
\begin{equation}\label{t }
\langle T_{strip}(w)\rangle=\left(\frac{\pi}{L}\right)^{2}
\sum_{j}\sum_{k}\frac{\lambda_{j}(\lambda_{k}+2Q\delta_{jk})z^{2}}
{(z-x_{j})(z-x_{k})}-\left(\frac{\pi}{L}\right)^{2}\frac{c}{24}.
\end{equation}
With this vacuum energy density, the Helmhlotz free energy, $F$,
could be computed to see how it changes if the length scale $L$ is
varied. This is a straightforward calculation \cite{difran} and
one finds that if a small variation $\delta L =\epsilon\,L$ is
applied to the strip width, the free energy varies in the
following way
\begin{equation}\label{free energy}
\delta F=\int\left(\frac{-\pi c}{24L^{2}}-\frac{\pi}
{L^{2}}\sum_{j,k}\lambda_{j}(\lambda_{k}+2Q\delta_{jk})\right)dw^{1}\delta
L.
\end{equation}
Integrating over $L$, the free energy of a SLE($\kappa,\rho)$
theory per unit length of the strip, $F_{u}$, is derived,
\begin{equation}\label{fin}
F_{u}=\frac{-\pi c}{24L}+\frac{-\pi}{
L}\sum_{j,k}\lambda_{j}(\lambda_{k}+2Q\delta_{jk}).
\end{equation}

The first term of the above equation is the usual part of free
energy in strip geometry, while the second term comes from the
presence of special boundary condition and reveals the specific
finite size properties of SLE($\kappa,\rho$). In the case where
$\sum\lambda_{k}+2Q=0$, boundary conditions do not change the free
energy. If the boundary changing operators be vertex operators of
coulomb gas, this condition which is the neutrality condition,
should be satisfied. That is, in the case which the boundary
operators are vertex operators, the free energy per length is
independent of boundary condition.

The whole procedure could be seen from a different point of view,
to derive central charge of some theories. Consider that the
charge $Q$ was not inserted in the coulomb gas action, so the
central charge would be one. Redoing all the above procedure one
arrives at the equation (\ref{fin}) with the charge term being
absent. Then the left hand side of this equation could be read as
$-\pi c'/24L$ with $c'$ being the effective central charge which
is read to be $c'=1+24(\sum\lambda_j)^2$. Thus if the sum of all
the charges in the theory be $2 Q=\sum\lambda_j$, then the new
central charge would be $c'=1+ 96 Q^2$. This argument is not
restricted to vertex operators and could be applied to any primary
field. If we use the ward identity of scale invariancy the result
is similar to the above: the change in free energy is zero if all
primary operators are in the finite distance, but sending one of
the operators to infinity, one finds that  the difference in the
free energy equal to $\frac{\pi h_{\lambda}}{L}$. The simplest
case is the ordinary SLE with one operator at origin and the other
at infinity. The change in free energy can be computed readily,
since both of the operators have equal weight
$h=\frac{6-\kappa}{2\kappa}$. A more complex case is SLE on
polygon \cite{Baufrie}, in this case we have
$\displaystyle{\sum\lambda_{j}=\frac{\sqrt{\kappa}}{2}}$ and the
free energy will be $\displaystyle{\frac{-\pi
c}{24L}-\frac{-\pi\sqrt{\kappa}}{2L}(\frac{\sqrt{\kappa}}{2}+2Q)}$.
This argument can also be applied to nSLE's \cite{dubda1}, in this
case $\rho_i=2$ and so
$\displaystyle{\sum\lambda_{i}=\frac{n}{\sqrt{\kappa}}}$ and the
free energy can be computed easily.

The equation (\ref{fin}) has some other important results in other
area, like quantum information theory. There is a relationship
between the free energy of a two dimensional classical statistical
system on a strip of infinite length and finite width $L$ and the
ground state energy of a one dimensional quantum system as
temperature $T\sim \frac{1}{k L}$, so instead of considering the
finite temperature quantum theory, one can use the equation
(\ref{fin}) to find the entropy of the corresponding quantum
system and then the specific heat of the model.

Additionally the method we have used to derive (\ref{fin}) is
applicable to quantum entanglement entropy. In the second method
of \cite{holzey} to find the quantum entanglement entropy of a one
dimensional quantum system on critical regime, one uses the
expectation value of energy-momentum tensor. Now we are able to
find the same quantity in presence of some additional boundary
operators. If the boundary operators be primary ones, then the
resulting entropy will not change. This comes directly from Ward
identities associated with scale invariancy, and so it's true for
arbitrary number of primary operators and also for more complex
CFTs like logarithmic CFTs. In these models we should use
different operator product expansions for the logarithmic
operators. Again sending the operators to infinity then the free
energy changes similar to equation (\ref{fin}) both for ordinary
CFT and LCFT.

The last point is that we are also able to find the behavior of
largest eigenvalue of the corresponding quantum system
$\ln\lambda_{0}=\frac{\pi c}{24L}+\frac{\pi}{L}\sum\lambda_{j}
\lambda_{k}+2Q\sum\lambda_{j}$ as $L$ tends to infinity. Note that
this is true if we are dealing with a unitary CFT, as in
nonunitary models some correlators may grow with distance.

\end{document}